# VEPP-2000 PROJECT

I. A. Koop, Budker Institute of Nuclear Physics, Novosibirsk, 630090, Russia

## Abstract

The status of VEPP-2M collider is presented. Implementation of Round Colliding Beams (RCB) concept in the new collider VEPP-2000 is outlined, potential advantages of RCB over the flat colliding beams are discussed. The main design parameters and features of this VEPP-2000 collider are reported.

## 1 MOTIVATIONS

Since the end of 1992 the $e^+e^-$ collider VEPP-2M in Novosibirsk has been successfully running in the c.m. energy range from threshold of hadron production up to 1.4 GeV. Since 1984 VEPP-2M is operating with the five poles superconducting wiggler with the maximum field $B = 8\,T$, which increases the beam emittance by a factor of 3. The integrated luminosity of about 50 pb$^{-1}$ was collected with two modern detectors SND[1] and CMD-2[2] allowing precise measurements of most of the hadronic channels of $e^+e^-$ annihilation. Together with 24 pb$^{-1}$ collected at VEPP-2M in the previous generation of experiments in 1974–1987, this integrated luminosity is more than one order of magnitude higher than about 6 pb$^{-1}$ accumulated by various experimental groups in Frascati and Orsay in the c.m. energy range from 1.4 to 2 GeV. Thus, there is a serious energy gap between the maximum energy attainable at VEPP-2M and 2 GeV in which existing data on $e^+e^-$ annihilation into hadrons are rather imprecise. Accurate measurements of hadronic cross sections in this energy range are crucial for better understanding of many phenomena in high energy physics.

A recent decision to upgrade the VEPP-2M complex by replacing the existing collider with a new one, in order to improve the luminosity and at the same time increase the maximum attainable energy up to 2 GeV, will significantly broaden the potential of experiments performed at the collider. Following modern modern trends, the new project was named VEPP-2000.

## 2 LUMINOSITY OF COLLIDERS AND ROUND COLLIDING BEAMS CONCEPT

The basic parameter of a collider is its luminosity $L$ which in the case of short bunches is determined by the formula:

$$L = \frac{\pi\gamma^2\xi_z\xi_x\varepsilon_x f}{r_e\beta_z}\cdot\left(1+\frac{\sigma_z}{\sigma_x}\right)^2,$$

where $\xi_x, \xi_z$ are the space charge parameters whose maximum values are limited by the beam-beam effects; $\varepsilon_x$ is the horizontal emittance of the beams, $\sigma_x, \sigma_z$ are their r.m.s. sizes at the interaction point (IP), and $\beta_z$ is the vertical $\beta$-function at the IP; $f$ is the frequency of collisions at this IP, $r_e$ is the classical electron radius, $\gamma$ is the relativistic factor.

The space charge parameter per interaction is:

$$\xi_{x,z} = \frac{Nr_e}{2\pi\gamma}\frac{\beta_{x,z}}{\left(\sigma_x+\sigma_z\right)\sigma_{x,z}},$$

where $N$ is the number of particles in the opposite bunch. Colliding bunches with maximum values of $\xi_z \approx 0.05$ and $\xi_x \approx 0.02$ are experimentally obtained on the VEPP-2M collider[3].

Aiming at a very high luminosity due to raising the $\xi$ limits in the Novosibirsk $\Phi$-Factory project[4, 5], colliding beams with round transverse cross-sections were proposed (just “round beams” in what follows). During the last decade at BINP, this approach evolved into the concept of Round Colliding Beams (RCB)[6].

In the RCB case, the luminosity formula has the form:

$$L = \frac{4\pi\gamma^2\xi^2\varepsilon f}{r_e^2\beta}$$

and the space charge parameters are now the same in the two directions, so the horizontal parameter can be strongly enhanced.

The evident advantage of round colliding beams is that with the fixed particle density, the tune shift from the opposite bunch becomes twice as small as the tune shift in the case of flat colliding beams. Besides, the linear beam-beam tune-shift in the round beams becomes independent of the longitudinal position in the bunch, thereby weakening the action of synchro-betatron resonances.

The main feature of the RCB is rotational symmetry of the kick from the round opposite beam; complemented with the $X-Z$ symmetry of the betatron transfer matrix between the collisions, it results in an additional integral of motion $M = xz' - zx'$, *i.e.* the longitudinal component of particle’s angular momentum is conserved. Thus, the transverse motion becomes equivalent to a one-dimensional (1D) motion. Resulting elimination of all betatron coupling resonances is of crucial importance, since they are believed to cause the beam lifetime degradation and blow-up.

The above arguments in favour of RCB have been checked out by the computer simulations of the beam-beam effects in the VEPP-2M collider lattice, modified to the RCB option[7]. The main results of the simulations[8] are presented in Fig. 1, where the beam sizes are plotted versus the space charge parameter $\xi$. One can see that the beam blow-up for the round beam option is much weaker than what is simulated by the same code for flat colliding beams (dashed line). The simulations have also demonstrated stability of RCB against the "flip-flop" effect, similarly to conclusions from simple flip-flop models[9].

## 3 DESIGN OF VEPP-2000

### *3.1 Magnets*

The lack of space for placing the new machine lead to demand on using of strong dipole magnets. For the energy of particles in the beam to be 1 GeV the field of a magnitude 2.3 T is needed. For this purpose it is planned to use the construction of magnets of booster BEP [10] that works at the level of the magnetic field needed. The parameters of magnets are shown in Table 1.

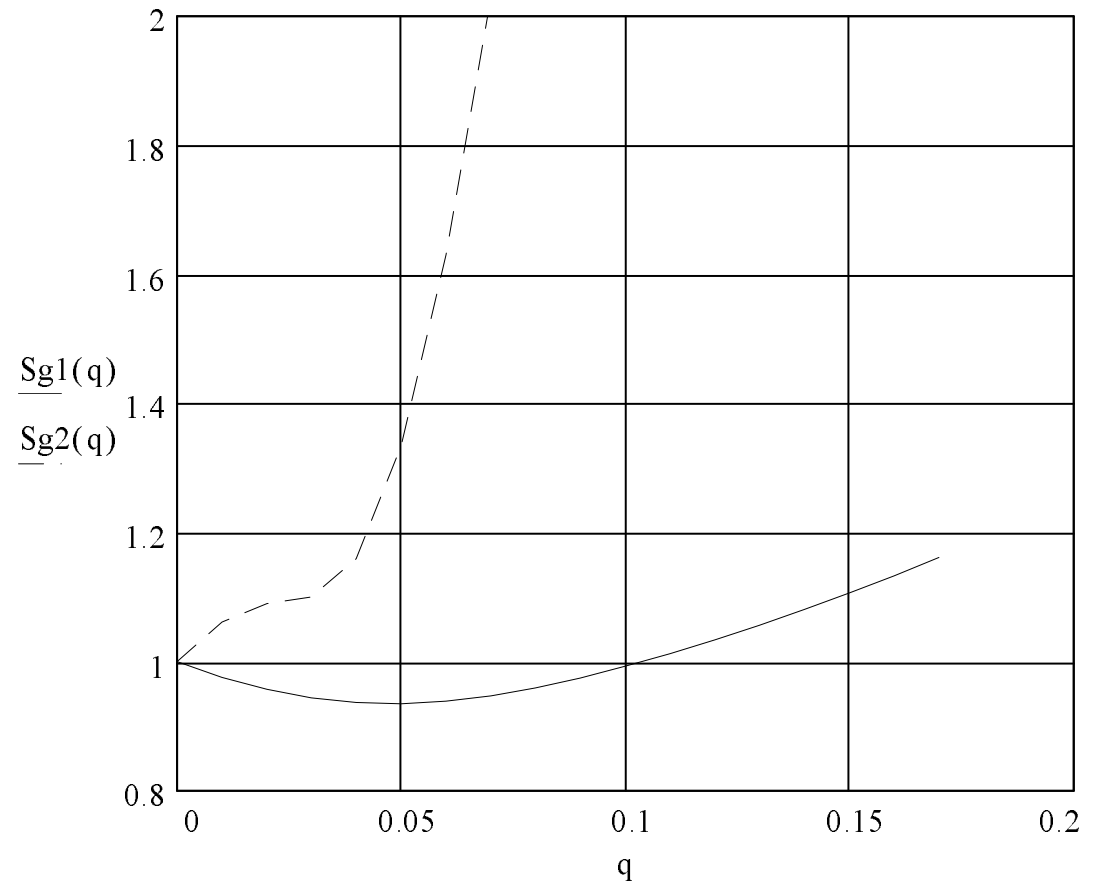

**Fig. 1 Variation of the weak beam ize vs. the space charge parameter $\xi$: solid curve for the round colliding beams, dashed curve for the flat beam option.**

**Table 1 The main parameters of the bending magnets**

| | |
|---|---|
| Gap | 40 mm |
| Bending angle | 45$^\circ$ |
| Bending radius | 1.40 m |
| Maximal field | 2.3 T |
| Number of coil turns | 10 |
| Current in coil | 9 kA |
| Power Consumption | 900 kW |
| Accuracy | 0.001% |
| Range of reconstruction with a beam | 1.2 |
| Time of reconstruction | 100 s |

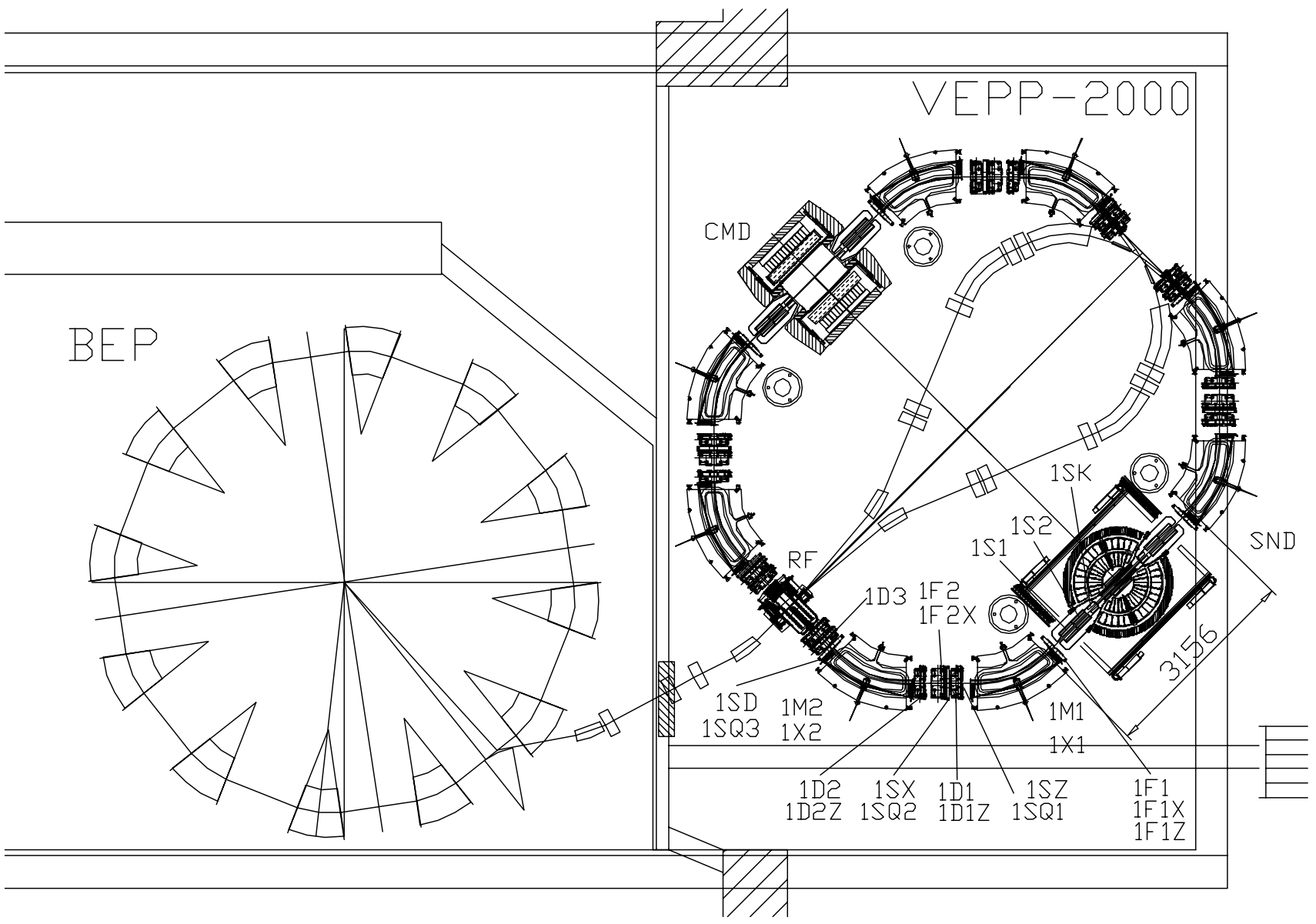

**Fig. 2 The VEPP-2000 collider layout**

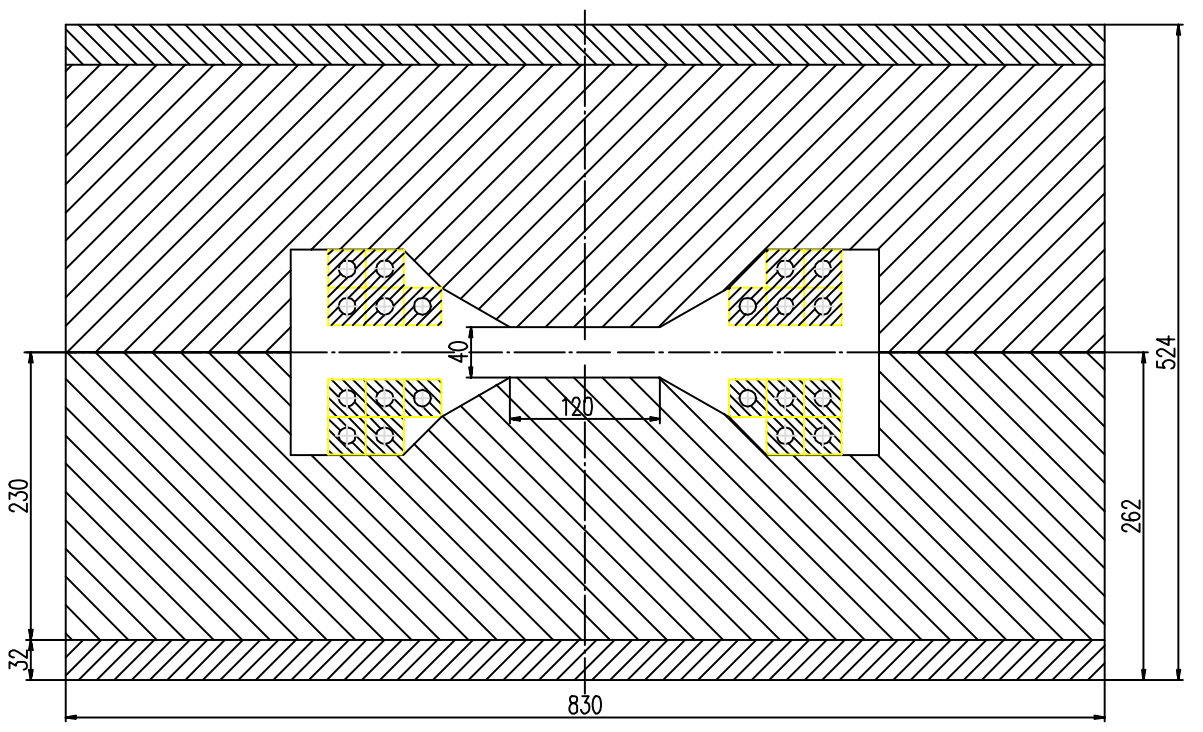

**Fig. 3 The dipole magnet of the VEPP-2000**

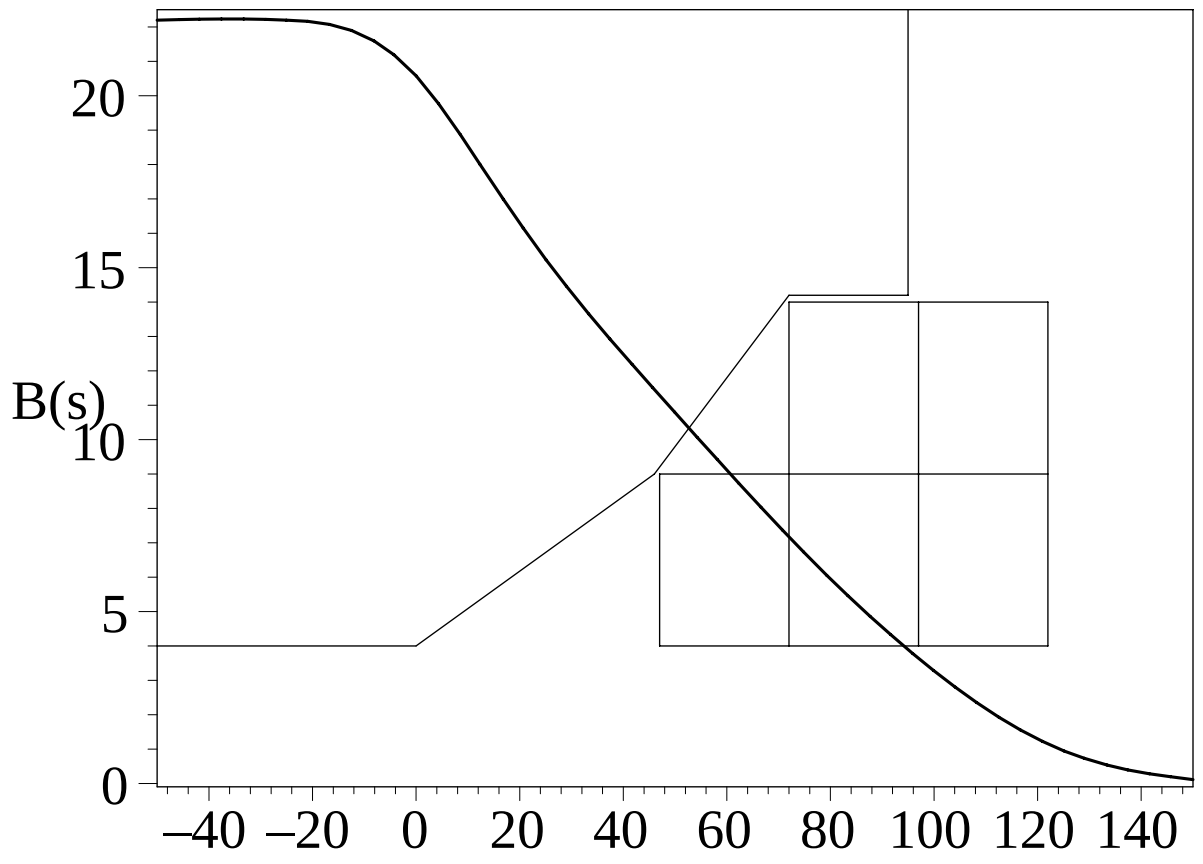

**Fig. 4 The dependence of the guide field of dipole magnet on the longitudinal coordinate (MERMAID). The edge of a magnet and its coil are shown schematically**

## *3.2 Solenoids*

Focusing in the two interaction regions is performed by SC solenoids, installed symmetrically with respect to the IPs. Each solenoidal block consists of a main solenoid which is longitudinally divided into two parts, and a compensating solenoid with reverse field to adjust longitudinal field integral and focussing. Such a scheme gives an additional possibility to control the *B* value by feeding only one half of the main solenoid at lower energies.

The solenoid coil is divided into three sections: inner section has thickness 30 mm and is made of $Nb_3Sn$ wire 1.23 mm in diameter (50% $Cu$ + 50% $Nb_3Sn$); middle section has thickness 20 mm and is wound with a $NbTi$ wire 1.2 mm in diameter (48% $Cu$ + 52% $NbTi$) and outer layer has thickness of 10 mm, made of $NbTi$ wire 0.85 mm in diameter (48% $Cu$ + 52% $NbTi$). To feed this three-section coil we plan to use two power supply units. Connection scheme implies that the current in the outer section is the sum of currents in the inner ones. The distribution of currents in the sections is: inner section - 145 A, intermediate section - 167 A, outer section - 312 A. The peak magnetic field is 12.1 T.

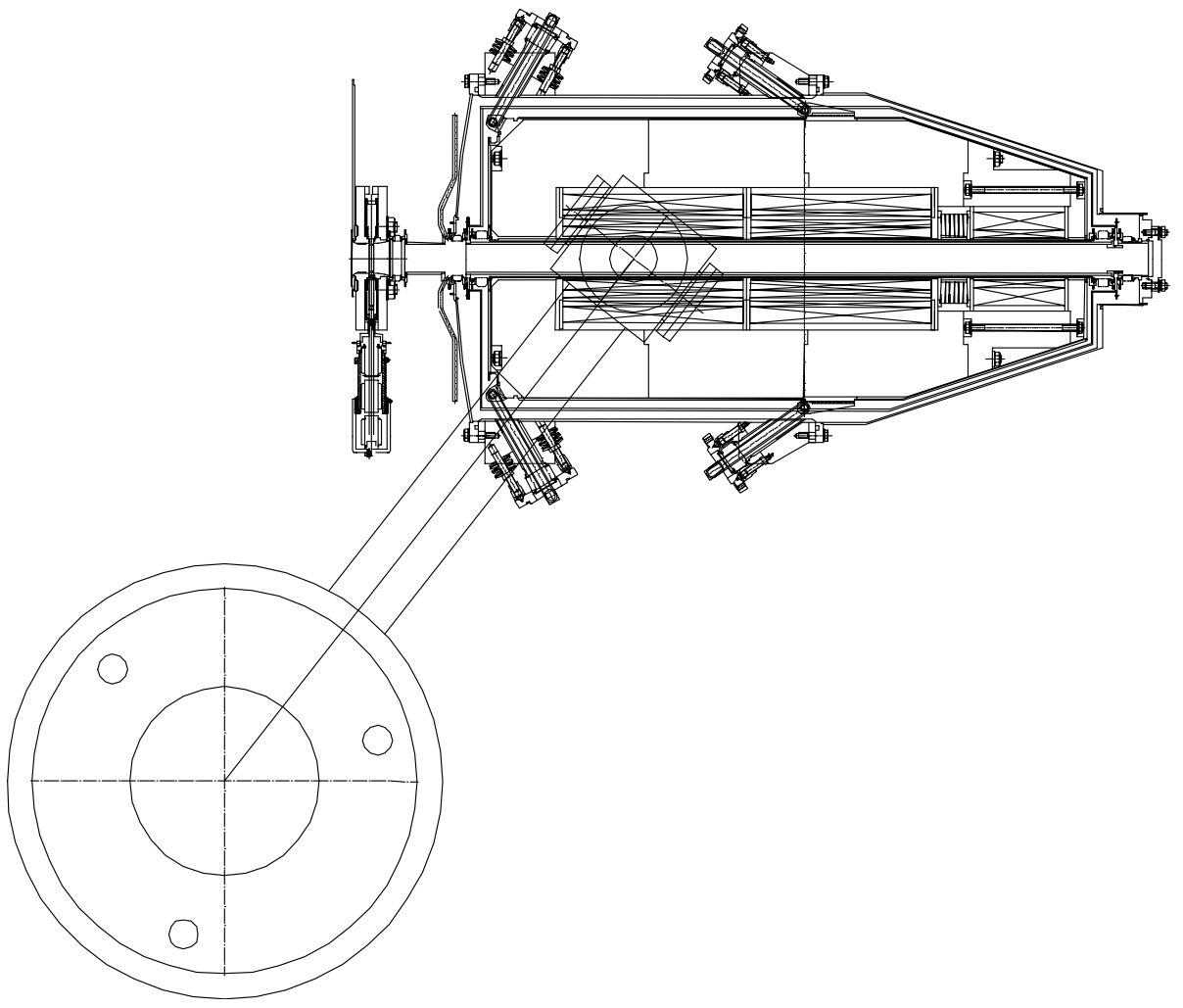

**Fig. 5 The superconducting solenoid of VEPP-2000**

**Table 2 The main parameters of solenoids**

| Solenoid | Main | Compensative |
|---|---|---|
| Magnetic field, T | 12.7 | 9.0 |
| Coil length, m | 0.526 | 0.128 |
| Inductance, H | 14.3 | 1.2 |
| Number of turns | 26080 | 4940 |
| Stored energy, J | 346 | 27 |

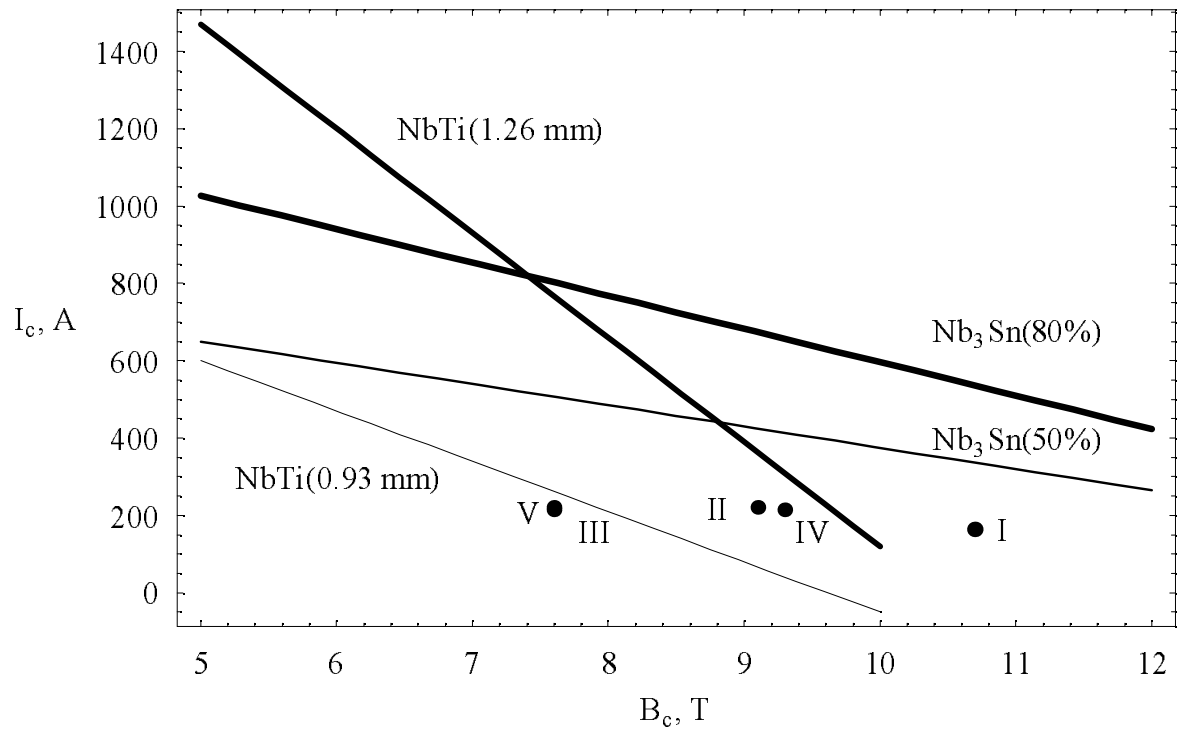

**Fig. 6 The critical parameters of the superconducting wires used. I — (10.7 T, 165 A), II — (9.1 T, 220 A), III — (7.6 T, 220 A), IV — (9.3 T, 214 A), V — (7.6 T, 214 A)**

Magnetic flux is closed by the iron return yoke located together with all the coils in a common LHe cryostat. Aperture of the coil is 50.0 mm. The inner tube of the helium vessel is a part of the collider vacuum chamber. A nitrogen vapour cooled liner is envisaged to protect the surface of the helium cryostat from heating by synchrotron radiation.

## *3.3 RF System*

Beam revolution frequency is 12.292 MHz. The accelerating RF frequency was chosen at 14-th revolution frequency harmonic i.e. 172 MHz. With accelerating voltage of 100 kV the bunch length is about $\sigma = 3$ cm at the energy of 1 GeV. Energy loss per turn is 64 keV, and with colliding beam currents of $2 \times 0.1$ A the power delivered to the beams is 12.8 kW. The so-called single-mode cavity is proposed to be used to ease suppression of coherent instabilities, see Fig. 7. Two coaxial damping loads are foreseen to absorb the energy from high-order modes excitation. The fundamental mode is isolated from the upper load by the tunable choke.

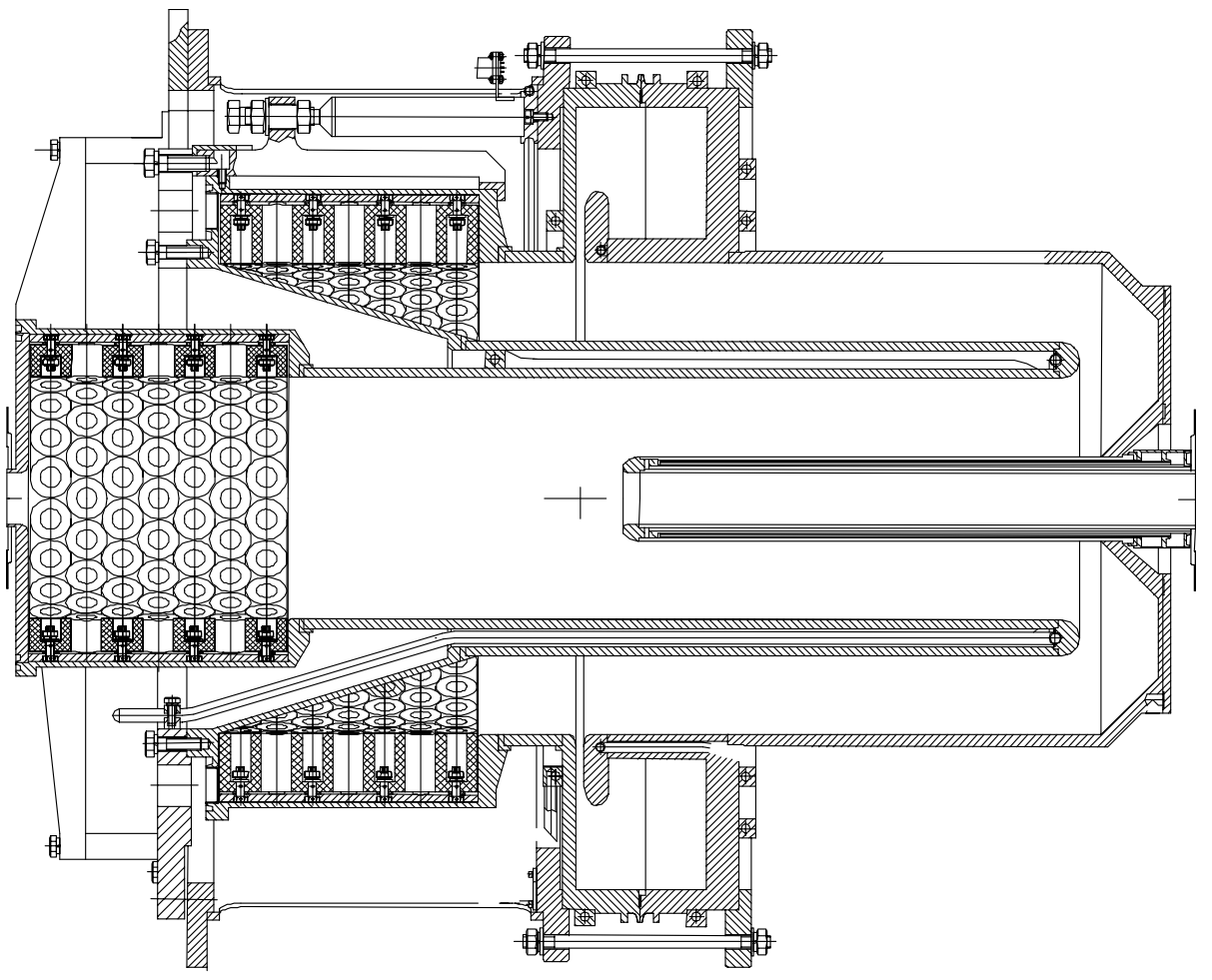

**Fig. 7 Cross-section of the cavity**

The RF field distribution of accelerating mode on the axis of the RF cavity is shown in Fig. 8.

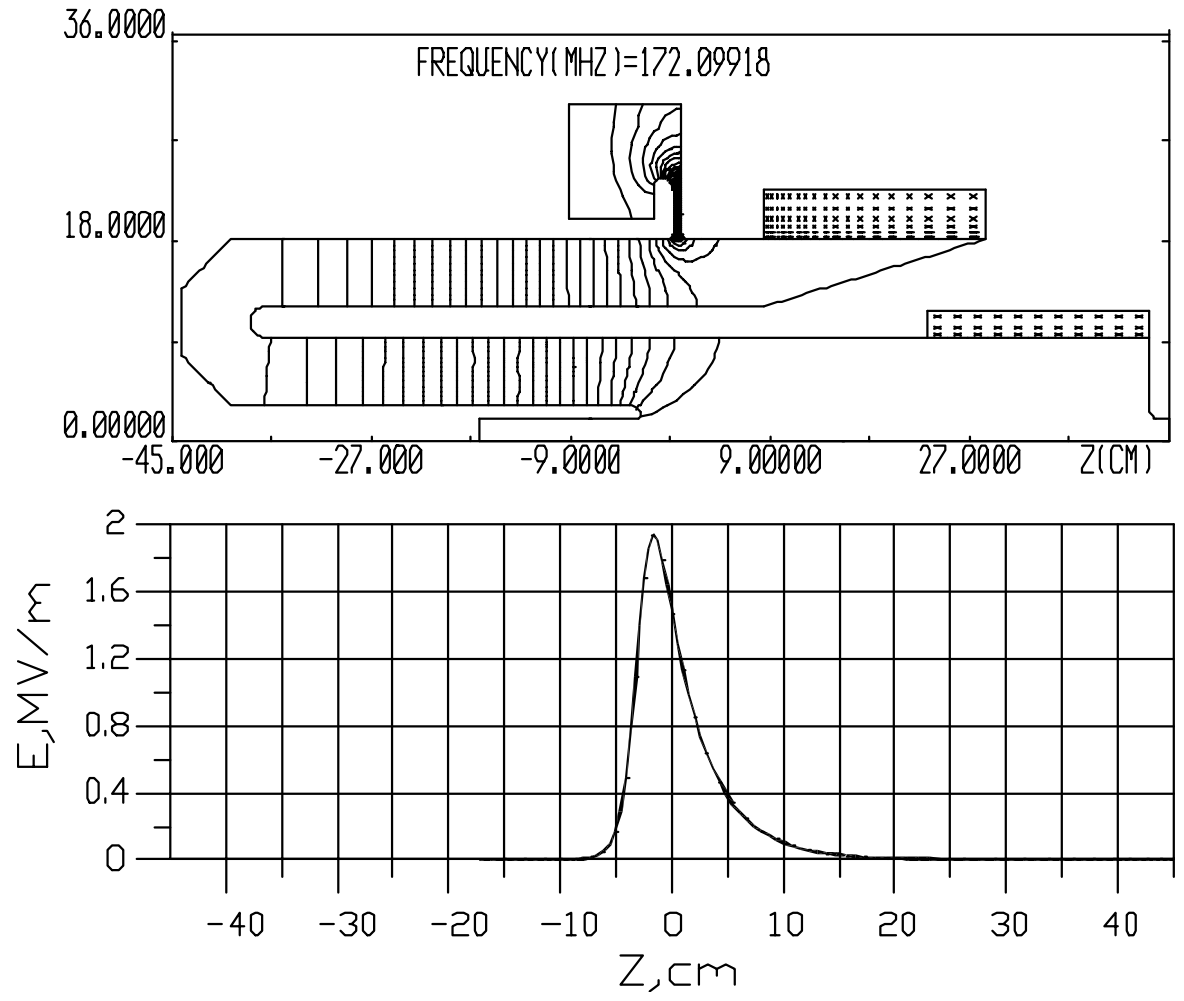

**Fig. 8 The RF field distribution of the accelerating mode of RF cavity (upper picture). The field distribution on the cavity axis (lower picture)**

## *3.4 Vacuum System*

The high-vacuum system consists of 16 ports with iongetter pumps PVIG-100, which are located at the edges of bending magnets vacuum chambers; of iongetter pump PVIG-250 connected to the RF cavity; of 4 cryopumps formed of cool solenoid surface. To prevent heating cryosurface, which is under 4.2 K, by SR there is planned to use perforated liner cooled with the liquid nitrogen. The liners ports should provide linear pumping rate of 5 l/s/cm for the nitrogen. The cool surface under 4.2 K is an ideal pumping for all the residual gases but the hydrogen, because after accumulation more than one cryosrpted hydrogen monolayer the saturated vapor pressure of the hydrogen reaches the value of $5 \cdot 10^{-7}$ Torr under given temperature. In spite of this circumstance our calculations showed that in general the beam life time will depend on residual CO pressure. Numerical simulation of the pressure over the ring was performed under following conditions:

- $Ip = Ie$ = 200 mA — electron and positron beam currents
- $S1, S2 \ldots S16$ — lumped ports with the pumping rate of 150 l/s, $Sp$ = 5 l/s/cm — distributed cryopumping
- Photon flux is $2.2 \cdot 10^{19}$ photon/s/rad
- Coefficient of photostimulated disorption (for after more than 100 Ah):
  - For non-heated chamber sections — $3 \cdot 10^{-5}$ molecule/photon

- For heated chamber sections — $3 \cdot 10^{-6}$ molecule/photon

The graph of CO pressure distribution for one quoter the VEPP-2000 circumference is presented below.

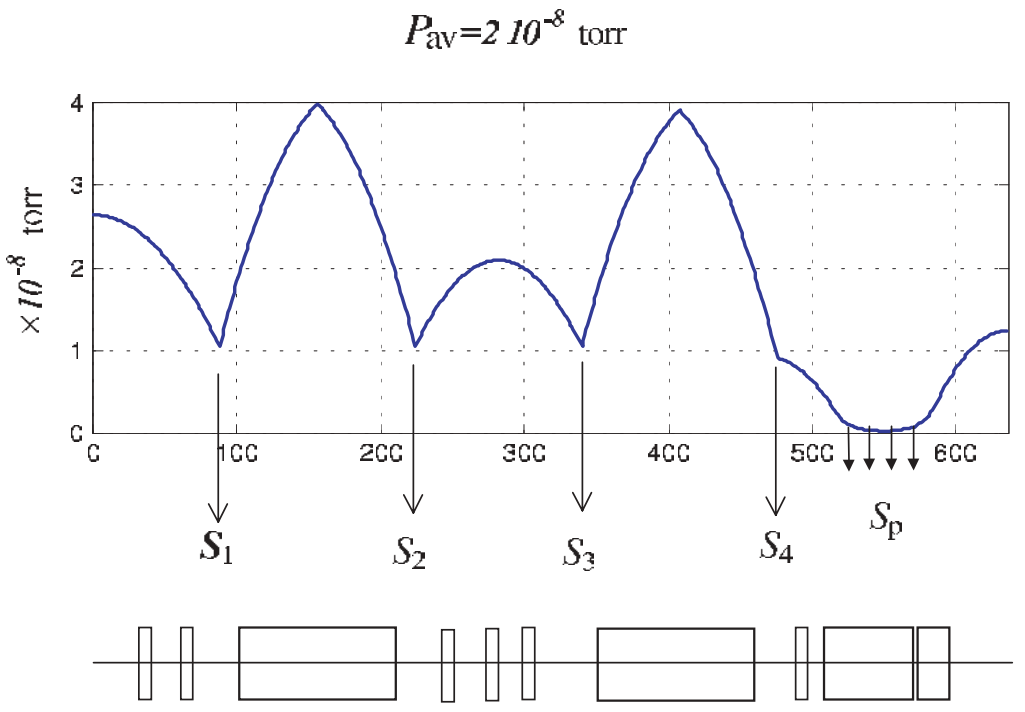

The vacuum chamber of the VEPP-2000 is shown in the Fig. 9.

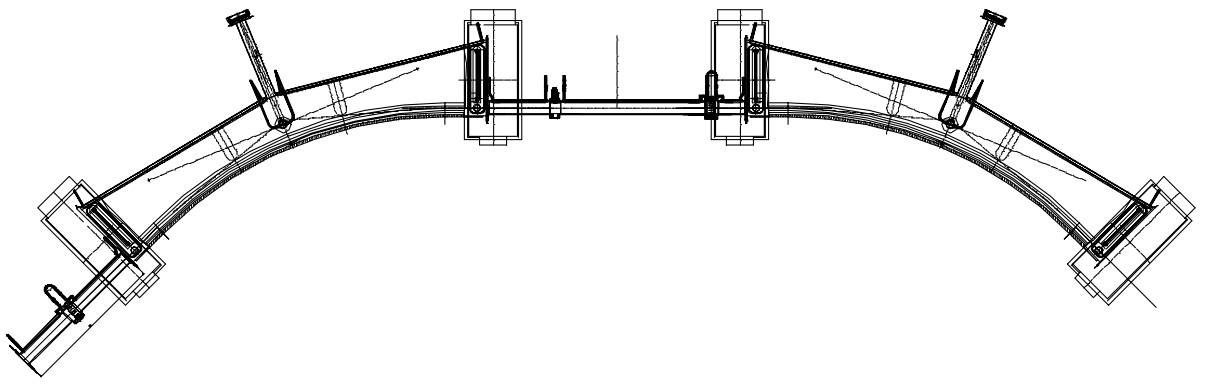

**Fig. 9 The vacuum chamber of one quadrant of VEPP-2000**

## 4 BEAM INJECTION

The injection of beams into the storage ring is planned to be done in the horizontal plane into the long drift opposite to the RF cavity (see the Fig. 2). The inflector plates will be placed on the inner side of the vacuum chamber in the bending magnets at the ends of the drift. The advantage of such a scheme is independence of the injected beam trajectory on the solenoids field. This gives us an opportunity to test different options of optics: usual round beams, "Möbius", and flat beams with zero rotation of the betatron oscillation plane.

In Table 3 the main parameters of generators are collected.

The BEP booster is capable of production beams with the energy of up to 900 MeV. Thus, operation at lower energies will be continuous, with injection of the beam at the experiment energy. In region from 900 MeV to 1 GeV the energy ramping from 900 MeV to the experiment energy is required.

**Table 3 The main parameters of kickers feed generators**

| Parameter | Value |
|---|---|
| Charging voltage | up to 40 kV |
| Polarity of charging voltage | positive |
| Amplitude of the output pulse | up to 30 kV |
| Half-height pulse duration | 80 ns |
| Pulse leading/trailing edge duration | 60 ns |
| Output pulse polarity | negative |
| Output impedance | 12.5 Ohm |
| Each circuite of DSC impedance | 6.25 Ohm |
| Thyratron current amplitude | up to 4.8 kA |
| Repetition frequency | 1 Hz |

## 5 VEPP-2000 PARAMETERS

The optical functions of the round beam lattice of VEPP-2000 are presented in the Fig. 10.

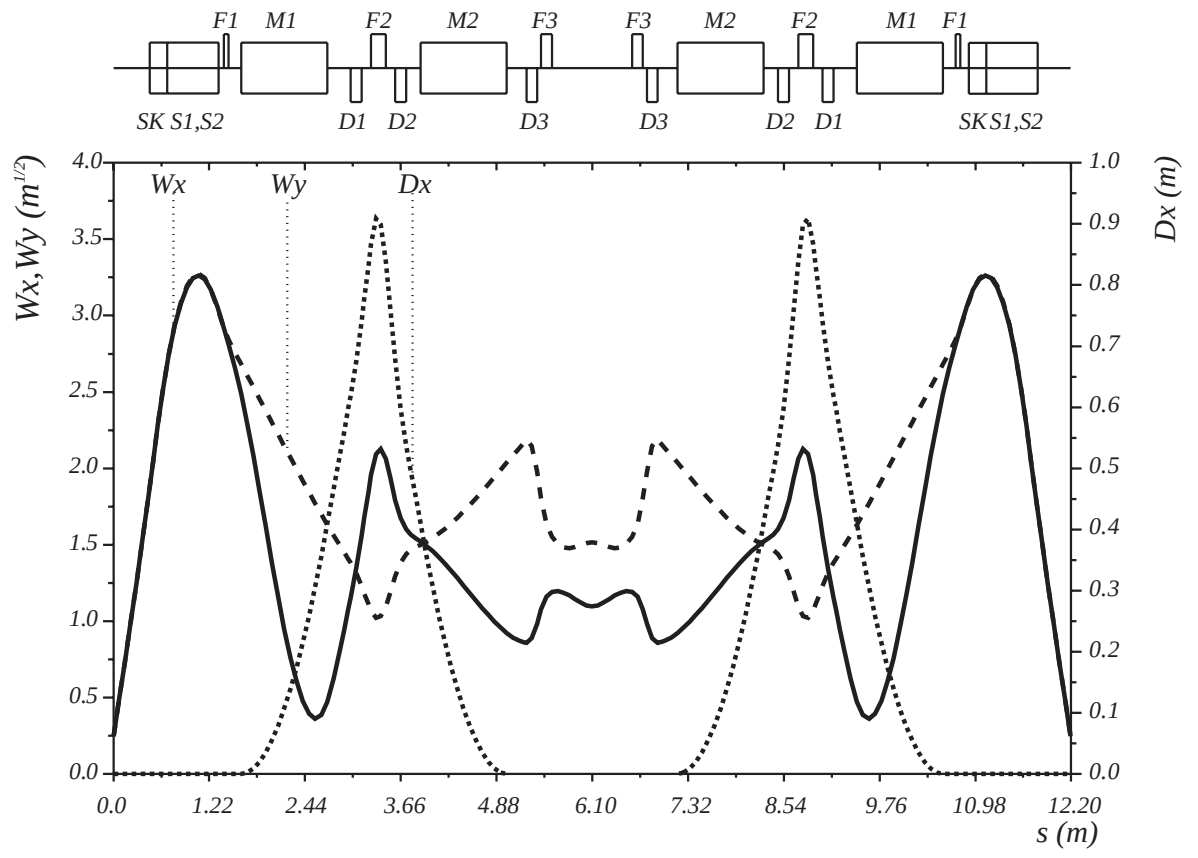

**Fig. 10 Half period lattice functions. $S = 0$ corresponds to IP**

An essential advantage of the found optics is zero dispersion in the IRs, RF cavity, and injection straight sections. The chosen optics has another very useful feature. Variation of the focusing strength of the solenoids changes $\beta_{IP}$ and the beam emittance in inverse proportion, at fixed energy. Changing energy, we can squeeze $\beta_{IP}$, conserving the maximum beam size at the solenoids, thus giving a possibility to tune optics for better performance. Apparently, this feature provides the luminosity scaling at lower energies approximately as $\gamma^2$ (instead of $\gamma^4$ for the option with fixed $\beta_{IP}$).

The main parameters of the new collider are given in Table 4.

**Table 4 Main parameters of the collider at E=900 MeV**

| | | |
|---|---|---|
| Circumference, m | C | 24.388 |
| RF frequency, MHz | $f_0$ | 172.0 |
| RF voltage, kV | V | 100 |
| RF harmonic number | q | 14 |
| Momentum compaction | $\alpha$ | 0.036 |
| Synchrotrone tune | $\nu_s$ | 0.003 |
| Emittances, cm·rad | $\varepsilon_x$<br>$\varepsilon_z$ | $2.2 \cdot 10^{-5}$<br>$2.2 \cdot 10^{-5}$ |
| Energy loss/turn, keV | $\Delta E_0$ | 41.5 |
| Dimensionless damping decrements | $\delta_z$<br>$\delta_x$<br>$\delta_s$ | $2.3 \cdot 10^{-5}$<br>$2.3 \cdot 10^{-5}$<br>$4.6 \cdot 10^{-5}$ |
| Energy spread | $\sigma_\varepsilon$ | $6.4 \cdot 10^{-4}$ |
| $\beta_x$ at IP, cm<br>$\beta_z$ at IP, cm | $\beta_x$<br>$\beta_z$ | 6.3<br>6.3 |
| Betatron tunes | $\nu_x, \nu_z$ | 4.1, 2.1 |
| Particles/bunch | $e^-, e^+$ | $1.0 \cdot 10^{11}$ |
| Bunches/beam | | 1 |
| Tune shifts | $\xi_x$<br>$\xi_z$ | 0.075<br>0.075 |
| Luminosity/IP, $cm^{-2} \cdot s^{-1}$ | $L_{max}$ | $1.0 \cdot 10^{-32}$ |

## 6 STATUS OF THE PROJECT

The designing of the main part of all the collider systems is finished. Manufacturing dipole magnets and quads is in progress and planned to be finished this year. Manufacturing and testing solenoids as the longest operation is planned to start this year and will continue the next year. The commissioning of the VEPP-2000 collider will start in the end of 2002.

## 7 REFERENCES

[1] M.N.Achasov et al., Preprint BudkerINP 98-65, Novosibirsk, 1998.

[2] R.R.Akhmetshin et al., Preprint BudkerINP 99-11, Novosibirsk, 1999.

[3] P.M.Ivanov *et al.*, in: *Proc. 3rd Advanced ICFA Beam Dynamics Workshop*, Novosibirsk, (1989), p. 26.

[4] L.M.Barkov *et al.*, in: *Proc. 14th Int. Conf. High Energy Accelerators*, Tsukuba (Japan), (1989), p.1385.

[5] L.M.Barkov et al., in: *Proc. of the IEEE Particle Accelerator Conf.*, San Francisco (1991), p.183

[6] V.V. Danilov et al., “The Concept of Round Colliding Beams”, in: *Proc. EPAC’96*, Barcelona (1996), p.1149.

[7] A.N. Filippov *et al.*, in: *Proc. 15th Int. Conf. High Energy Accelerators*, Hamburg (Germany), (1992), p.1145.

[8] I. Nesterenko *et al.*, in: *Proc. 1997 PAC*, Vancouver (Canada), (1997), p. 1762.

[9] A.V. Otboyev and E.A. Perevedentsev, in: *Proc. 1999 PAC*, New York (1999).

[10] V.V.Anashin et al., Preprint BudkerINP 84-114, Novosibirsk, 1984.